\documentclass{emulateapj}
\usepackage{amsmath,amstext,amssymb}
\usepackage{hyperref}

\usepackage{graphicx}

\slugcomment{The Astrophysical Journal, 748:145, 2012 March 16}
\shorttitle{NGC 1097: double-peaked H$\alpha$ variability}
\shortauthors{Schimoia et al.}

\begin{document}

\title{Short timescale variations of the H$\alpha$ double-peaked profile of the nucleus of NGC\,1097}

\author{Jaderson S. Schimoia$^{1}$}
\author{Thaisa Storchi-Bergmann$^{1}$}
\author{Rodrigo S. Nemmen$^{2}$}
\author{Cl\'audia Winge$^{3}$}
\author{Michael Eracleous$^{4}$}

\affil{$^1$Instituto de F\'isica, Universidade Federal do Rio Grande do Sul, Campus do Vale, Porto Alegre, RS, Brazil; silva.schimoia@ufrgs.br}
\affil{$^2$NASA Goddard Space Flight Center, 8800 Greenbelt Road, Greenbelt, Maryland, USA}
\affil{$^3$Gemini South Observatory, c/o AURA Inc., Casilla 603, La Serena, Chile}
\affil{$^4$Department of Astronomy and Astrophysics, the Pennsylvania State University, 525 Davey Lab, University Park, PA 16802, USA}

 \begin{abstract}

The broad (FWHM $\sim$ 10,000 km s$^{-1}$)
double-peaked H$\alpha$ profile from the LINER/Seyfert 1 nucleus of NGC\,1097
was discovered in 1991, and monitored for
the following 11 years. The profile showed variations attributed to the rotation of gas in a
non-axisymmetric Keplerian accretion disk, ionized by a varying radiatively inefficient accretion flow (RIAF) 
located in the inner parts of the disk.
We present and model 11 new spectroscopic
observations of the double-peaked profile taken between 2010 March and 2011 March. This series of observations was 
motivated by the finding that in 2010 March the flux in the double-peaked line was again strong, becoming,
in 2010 December, even stronger than in the observations of a decade ago.
We also discovered shorter timescale variations than in the previous observations:
(1) the first, of $\sim$7 days, is interpreted as due to reverberation'' of the
variation of the ionizing source luminosity, and the timescale of 7 days as the 
light crossing time between the source and the accretion disk; this new timescale and its interpretation provides a  distance between the emitting gas and the supermassive black hole and as such introduces
a new constraint on its mass; (2) the
second, of $\approx$ 5 months, was attributed to the rotation of a
spiral arm in the disk, which was found to occur on the dynamical timescale. We use two accretion disk
models to fit theoretical profiles to the new data, both having non-axisymmetric
emissivities produced by the presence of an one-armed spiral. Our
modeling constrains the rotation period for the spiral to be
$\approx$\,18 months. This work supports our previous conclusion that
the broad double-peaked Balmer emission lines in NGC\,1097, and
probably also in other low-luminosity active nuclei, originate from an accretion disk
ionized by a central RIAF.

 \end{abstract}


\section{Introduction}

The barred spiral galaxy NGC\,1097 has an active nucleus classified as
LINER (Low-Ionization Nuclear Emission-Line Region, \citet{Heckman}), whose spectrum
was found in 1991 to show transient broad
(FWHM\,$\approx$\,10,000\,km\,s$^{-1}$) double-peaked Balmer lines
\citep{SB93}. The nucleus of NGC\,1097 was the first low-luminosity active galactic nucleus (LLAGN) in a LINER 
discovered displaying such lines, and, because of these broad lines, has also been 
classified as a Seyfert\,1 nucleus. Later, other LLAGNs in LINER nuclei were found to display 
double-peaked Balmer lines, such as
M\,81 \citep{Bower}, NGC 4203 \citep{Shields}, and NGC 4450 \citep{Ho}.  

The double-peaked H$\alpha$ profile of the NGC\,1097 nucleus
was monitored for the following 11 years after the discovery \citep{SB03} (SB03), showing variability in its
flux, width and relative intensity of the blue and red peaks. These variations were
attributed to gas emission from a thin Keplerian accretion disk, with a
non-axisymmetric perturbation around a
supermassive black hole (hereafter SMBH), where the gas is ionized by a central source,
\citep{Eracleous95, SB95, SB97, SB03}.

The variability of the double-peaked profile of NGC\,1097 is not
rare. The first active nucleus found to display variable double-peaked broad Balmer emission lines was the
nucleus of the radio galaxy 3C\,390.3 \citep{Yee}. This object can be considered prototypical of the class of double-peaked emitters, because many of the now well-known features of these objects, such as variability in the relative intensity of the blue and red peaks \citep{Veilleux, Zheng}, were first observed in 3C\,390.3. Another remarkably well studied double-peaked emitter is the active nucleus of
Arp\,102B. For this object, a study by  \citet{CeH89} has
shown that  the double-peaked Balmer lines cannot
be driven by local viscous dissipation in the line-emitting part of
the disk, as the H$\alpha$ luminosity is of the order or even exceeds the 
energy locally available to power the line. The same conclusion was obtained for other double-peaked
emitters by \citet{Eracleous94}  and \citet{Strateva06, Strateva08}. These authors suggested that the lines are powered by illumination from an external source. The ionizing source could be an
geometrically thick accretion flow, located within the inner radius of the line-emitting
portion of the disk.  In the case of NGC\,1097, \citet{Nemmen06} found
that the spectral energy distribution (SED) of the nucleus is indeed well
described by such a structure, a radiatively inefficient accretion
flow (RIAF, \citet{Narayan08}). This has led to the conclusion that an
inner RIAF is the ionizing source of the accretion disk in NGC\,1097.

\citet{Lewis10} and \citet{Gezari} reported a long-term monitoring of 14 other
double-peaked emitters and found similar variations to those observed for NGC\,1097,
3C390.3 and Arp\,102B.  These objects also display in at least one epoch the red
peak more intense than the blue peak. This indicates that most
line-emitting disks are non-axisymmetric, as in an axisymmetric disk
the blue peak should be always more intense than the red because
of Doppler boosting. The disks could then be elliptical
\citep{Eracleous95}, or have a non-axisymmetric emitting structure,
such as a hot spot or spiral arm \citep{Lewis10}.

Another model which can reproduce double-peaked emission-line profiles is the 
one proposed by \citet{GKN} and \citet{Gaskell10}, in which the 
ionizing source is the continuum from a thin accretion disk, and the origin of the 
double-peaked lines is the broad-line region (BLR) which is proposed to be an inner 
extension of the torus, in the shape of a ``bird's nest''.  This geometry is not that 
distinct from that of the outer parts of an accretion disk,  and in this regard is 
in approximate agreement with our proposed geometry. But, in the case of NGC\,1097, 
the study by \citet{Nemmen06} favors an RIAF origin for the nuclear ionizing continuum 
instead of a hot and thin accretion disk.

The monitoring of double-peaked profiles is a valuable experiment in
the investigation of the structure and dynamics of the accretion disks
and how they evolve as the gas moves inwards to feed the SMBH. In
\citet{SB03}, observing the NGC\,1097 H$\alpha$ double-peaked profile
approximately once or twice a year, we concluded that the asymmetry of the
double-peaked profile was due to a spiral arm perturbation in the disk, 
rotating with a period of 5.5 years. But as relevant timescales 
for evolution of the innermost regions of AGNs
can be as short as weeks and months, we decided to repeat our
experiment on NGC\,1097 in order to probe shorter timescales. Here we
report the results of this experiment: we indeed found short variability
timescales. We also verified that, although our last observations of
2001 showed that the profile was becoming fainter and fainter, it is
now again strong and was even stronger in 2010 December than in
all previous observations.

This paper is organized as follows: in Section 2 we describe the
observations and the data reduction in Section 3 we present the observational results, in
Section 4 we describe the models and the results from the modeling. In Section 5 we
discuss the timescales of the accretion disk variability and the
interpretation of the modeling. The conclusions of this work are
presented in Section 6.

\section{Observations and Data Reduction}
\label{observations}

Long-slit optical spectra of the nucleus of NGC 1097 were
obtained in queue mode in 11 epochs from 2010 March 4 to 2011 March 21 using the Gemini South Multi-Object Spectrograph (GMOS)
in the long-slit mode.
The allocation was part of the ``Poor Weather''
opportunity, which is ideal for spectral monitoring which does not
need absolute flux calibration or regular cadence. This is the case
for the nucleus of NGC\,1097, for which we use the integrated flux in
the narrow emission lines to normalize the spectra to match previous
observations, under the assumption that the narrow lines do not vary
within the extraction aperture.

 The first three observations were obtained at time intervals of 4-5 months, since 
our previous monitoring suggested 
significant variations on a timescale of years.
However, our new observations showed strong variations in the nuclear
spectrum on timescales of months, which motivated us to ask for new
observations at shorter time intervals. We were then able to get at
least two observations a month from 2011 January to 2011 March, as
shown in the observing log presented in Table \ref{obslog}. This table
lists the identification of the program, date of observation, position angle
(p.a.) of the slit and the average seeing during the observations,
measured as the FWHM of a spatial profile extracted along a range of
columns containing only the broad line (after subtraction of the
underlying continuum).
 
The data were obtained with the GMOS South with a slit width of
1$\farcs$0 ($\sim 80\,$pc), using the B600 grating and the GG455 filter to block
possible higher order contamination. The spectral range covered was
6000--7000\AA\, for most spectra, at a spectral resolution of $\sim$
4.5\AA \,(FWHM of the lines in the arc spectrum, $\approx 200$ km s$^{-1}$). Six 600s exposures
were obtained at each epoch, which were then combined into a final
spectrum.


\begin{deluxetable}{l l l l l}
\centering
\tablecolumns{5}
\tablecaption{Observation Log \label{obslog}}
\tablehead{
			& 			& p.a.			& Seeing \\
 Program ID		&UT Date 		&(deg)			&(arcsec)}
\startdata
GS-2009B-Q-99		& 2010 Mar 4		& 70			&1.00	\\
GS-2010A-Q-81		& 2010 Aug 15		& 55			&0.98	\\
GS-2010B-Q-90		& 2010 Dec 22		& 280			&0.63	\\
GS-2010B-Q-90		& 2011 Jan 11		& 280			&0.73	\\
GS-2010B-Q-90		& 2011 Jan 15		& 280			&0.73	\\
GS-2010B-Q-90		& 2011 Jan 18		& 280			&0.70	\\
GS-2010B-Q-90		& 2011 Feb 8		& 280			&1.21	\\
GS-2010B-Q-90		& 2011 Feb 10		& 280			&0.79	\\
GS-2010B-Q-90		& 2011 Feb 14		& 280			&0.72	\\
GS-2010B-Q-90		& 2011 Mar 17		& 280			&1.43	\\
GS-2010B-Q-90		& 2011 Mar 21		& 280			&0.82
\enddata
\end{deluxetable}

The slit width of 1$\farcs$0 was selected because it is usually larger
than the average seeing during the observations and was also the slit
width of previous observations (SB03). Using the same slit width
allows us to normalize the new spectra so that they match the flux of
the narrow emission lines observed in previous spectra and study the
evolution of the flux and profile of the broad H$\alpha$ line over
$\approx$\,20\,yr.

The data reduction was performed using the standard tasks from the
{\it gmos} package in the Image Reduction and Analysis
  Facility software (IRAF\footnote{IRAF is distributed by the National Optical Astronomy Observatory, 
which is operated by the Association of Universities for Research 
in Astronomy, Inc., under cooperative agreement with the National 
Science Foundation.}). We extracted the nuclear spectra
using a window of 1$\farcs$0 $\times$1$\farcs$0 centered at the peak
of the continuum emission, assumed to correspond to the galaxy
nucleus. This position coincides also with the location of the
unresolved source of the double-peaked H$\alpha$ lines.
Figure \ref{acquisition} shows the acquisition image of the
observation of 2010 August 15, where the nuclear extraction region is
represented by the square labeled A.

\subsection{Stellar population contribution}
\label{stellarpopulation}

\begin{figure}
 \centering
  \includegraphics[scale=0.5]{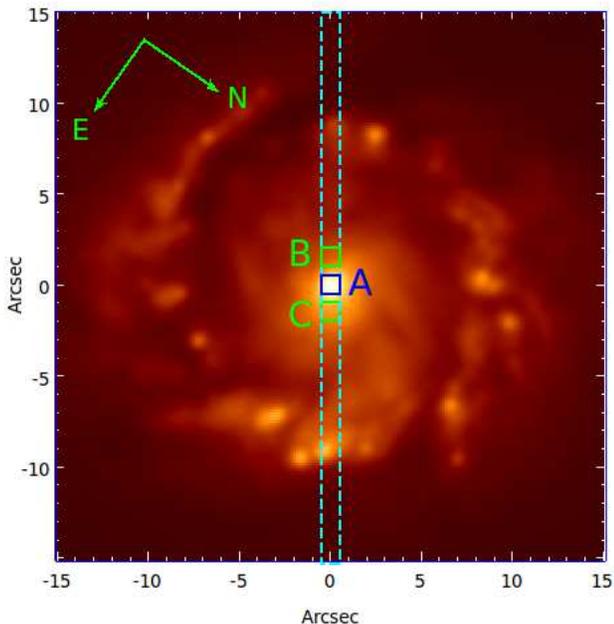}
  \caption{Acquisition image from 2010 August 15, where we illustrate
    the locations of the 1$\farcs$0$\times$1$\farcs$0 extraction
    windows for the nucleus and stellar population spectra. The square
    labeled A represents the nuclear extraction window, while squares
    B and C represent the extranuclear ones, from which we obtained
    the spectrum of the underlying stellar population.}
  \label{acquisition}
\end{figure}

Since the nuclear spectra show strong absorption lines from the
underlying stellar population, we subtracted the stellar population
contribution, in order to isolate and better
study the emission line profile and in particular the broad H$\alpha$ line. A
template stellar population spectrum was obtained as follows. For each
epoch, in addition to the nuclear spectrum, we extracted two more
spectra using two similar windows of 1$\farcs$0\,$\times$\,1$\farcs$0,
one centered at 1$\farcs$5 to the southwest (Figure \ref{acquisition},
region B) and another at 1$\farcs$5 to the northeast of the nucleus
(Figure \ref{acquisition}, region C). These extranuclear spectra do
not show any broad H$\alpha$ emission, but show some narrow-line
emission which was edited out, using as guides, stellar population
templates from previous studies \citep{Bica}. Averaging these two
spectra, we obtain a stellar population spectrum from a location
approximately 100\,pc away from the nucleus. Under the assumption that
this spectrum is representative of the stellar population at the
nucleus, we subtracted it from the nuclear spectrum (after
normalizing the continuum to the corresponding value at the nucleus)
thus isolating the gas emission.

\begin{figure}
 \centering
  \includegraphics[scale=0.55]{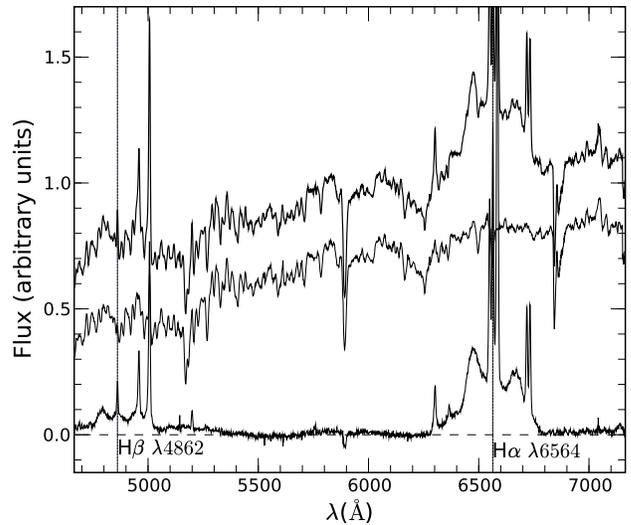}
  \caption{Observation from 2010 August 15. Top: nuclear spectrum
    (region A of Figure \ref{acquisition}). Middle: stellar population
    obtained from averaging the spectra from regions B and C in Figure \ref{acquisition}. Bottom: nuclear spectrum after the
    subtraction of the stellar population.}
  \label{exemp1}
\end{figure}

 Figure \ref{exemp1} illustrates the
process for the spectrum of 2010 August 15, which covers a broader
spectral range than most spectra, from 4750 to 7200\AA.
We note that this process reveals the presence of a broad double-peaked H$\beta$ line
which was not obvious prior to the subtraction, as well as some weak emission lines
such as [N\,{\sc i}]\,$\lambda$5199\AA\,and [O\,{\sc i}]\,$\lambda$6363\AA.
The Na\,I\,D absorption line in the residual spectrum is most probably interstellar, as all other
stellar absorption lines did not show any residual.

We also note that after subtracting the stellar population spectrum from the nuclear one,
we do not see any evidence for a significant contribution of an AGN continuum, as can be seen in Figure \ref{exemp1}. The effect of a nuclear continuum, if present, would be to dilute the stellar absorption lines, making them shallower at the nucleus than outside. The subtraction of the stellar population template, as described above, after normalization to the flux in the continuum of the nuclear spectrum  would result in the presence of ``fake'' emission features in the residual spectrum and a possible slope (in the case of a blue continuum, for example), which are not seen.

We have performed a sequence of  tests in order to estimate the contribution from a featureless continuum (FC) that could be ``hidden'' in the nuclear spectrum as follows. We added the contribution of the FC, assumed in the form  $F_{\nu} \propto \nu^{1.5}$, to the stellar population template described above, at increasing percent contributions of its flux at 5800\AA. The resulting composite (FC+stellar population) spectrum was then renormalized to the same flux at 5800\AA, and we subtracted the stellar population template from it. We concluded that for the short wavelength range and the signal-to-noise ratio of our observations, in order to have clear residuals showing fake emission features and a blue continuum we need the FC continuum to contribute with at least 15\%--20\% of the flux at 5800\AA. We thus can say that a possible FC contribution to the nuclear spectrum is  lower than 20\% of the continuum flux at 5800\AA.
We consider this value an upper limit, as a comparison between the stellar continuum from our spectra and the AGN continuum obtained in the work of Nemmen et al. (2006) shows that the stellar continuum in our aperture is about 30 times stronger than the AGN continuum in the optical.

\section{Results}

After subtracting the contribution of the stellar population, the
final step before the analysis of the double-peaked profile was to
normalize the spectra using as reference a previous flux-calibrated
spectrum. We used as reference the spectrum of 1991 November 2
\citep{SB93}, and assumed that the flux of the narrow lines H$\alpha$,
[N\,{\sc ii}]\,$\lambda\,\lambda$6548, 6584, and [S\,{\sc ii}]\,$\lambda
\lambda$6717, 6731 did not vary during the last 20 years. This
is a common assumption in this type of study, as the narrow
lines originate in extended regions with spatial scales much larger ($\sim$ 100\,pc) than that of the
accretion disk ($\sim\,10^{-3}$pc). Thus, their response time to variations in the ionizing flux is 
correspondingly much larger than that for the line-emitting accretion disk,
and cannot respond  to variations of the ionizing flux that are as fast as those
we have detected. 

\begin{figure}
 \centering
  \includegraphics[scale=0.55]{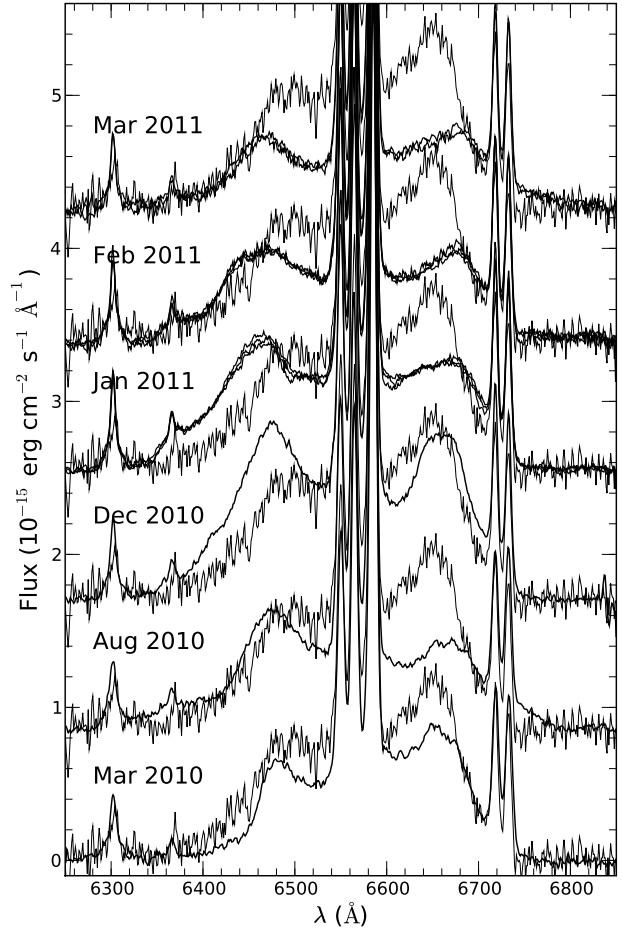}
  \caption{Nuclear spectra from 2010 March to 2011 March each compared with
    the reference spectrum of 1991 November 2 (gray).  For 2011 January--March we show the spectra obtained a few days
    apart plotted on top of each other, as no significant changes were
    found between them.}
  \label{results}
\end{figure}

The normalized spectra from our 2010--11 observations are compared with
the reference spectrum of 1991 November in Figure\,\ref{results} (pairs
of spectra are shifted from each other in the vertical direction for
clarity). In this figure we show the three spectra from 2011 January,
the three from 2011 February, and the two from 2011 March,
respectively, on top of each other because we did not find any
significant variation within each group.

Note that the double-peaked H$\alpha$ profile from 2010 August is very
different from that of 2010 March. In 2010 March, the red peak was
stronger than the blue peak, while in 2010 August, the blue
peak was stronger than the red. This inversion of the relative
intensity of the peaks occurred on a timescale of 5 months. In
addition, in 2010 December, the profile showed a marked increase in
its integrated flux compared to that of 2010 August, while the
red and blue peaks showed similar strengths. 

From 2010 December 22 to 2011 January 11 we observed a $\sim$\,25\%
decrease in the integrated flux of the double-peaked profile. The
three observations of 2011 January were obtained within a time interval
of seven days, and showed no significant variations.

In 2011 February, the three observations were obtained within a time
interval of six days, and again we did not see significant variations
among the three profiles. However, the total flux continued to
decrease compared with that of January, with the blue and red
peaks showing similar strengths.

In 2011 March, we obtained two observations separated by a time
interval of four days, between which we again did not find significant
variation in the double-peaked profile. But the total flux decreased
further, and the profile showed the red peak somewhat stronger
than the blue peak.

In summary, this series of observations obtained between 2010 March
and 2011 March reveal: (1) a marked increase of the broad-line flux
between 2010 August and 2010 December; (2) a significant decrease of the
flux on timescales longer than a week; (3) no significant changes in
the profile on time intervals shorter than a week; (4) an inversion in
the relative strength of the blue and red peaks on a timescale of
5 months.

\subsection{Measurements of the double-peaked profile characteristics}

We have quantified the observed variations by measuring the
wavelengths of the blue and red peaks of the profiles, their peak flux
densities, and the integrated flux of the broad line, as described
below. We list the results in Table \ref{medidas}.

In order to isolate the broad double-peaked H$\alpha$ emission and
perform the measurements, we first evaluated the contribution of the
narrow lines H$\alpha_{\text{narrow}}\,\lambda6564$, [N\,{\sc ii}]\,$\lambda\lambda6549, 6584$, 
[S\,{\sc ii}]\,$\lambda\lambda6718, 6732$ and
[O\,{\sc i}]\,$\lambda\lambda6302, 6365$ by fitting gaussians to these lines
and then subtracting them.  The flux of the double-peaked line ($F_{\text{broad}}$) was
measured by integrating the total flux under the profile after this
subtraction. The average uncertainty in $F_{\text{broad}}$ is
$1\,\times\,10^{-14}$ erg cm$^{-2}$ s$^{-1}$.

The wavelengths of the blue and red peaks ($\lambda_{B}$ and
$\lambda_{R}$) and the corresponding peak flux densities ($F_{B}$
and $F_{R}$) were obtained by fitting three Gaussians to the
double-peaked profile, one for the blue peak, another for the red peak
and a third Gaussian for the center of the profile. $F_{B}$ and
$F_{R}$ are the peak flux densities of the Gaussians fitted to the
blue and red peaks, respectively, while $\lambda_{B}$ and
$\lambda_{R}$ are the corresponding wavelengths. We have tried other
methods as well, such as using splines to fit the peaks, but concluded
that the fit of Gaussians gave equivalent but more robust results. The
average uncertainties in the measurements are $\approx 4$\AA\,for
$\lambda_{B}$ and $\lambda_{R}$ and $\approx 6\,\times\,10^{-17}$ erg
cm$^{-2}$ s$^{-1}$\AA$^{-1}$ for $F_{B}$ and $F_{R}$.

\begin{figure}
 \centering
  \includegraphics[scale=0.55]{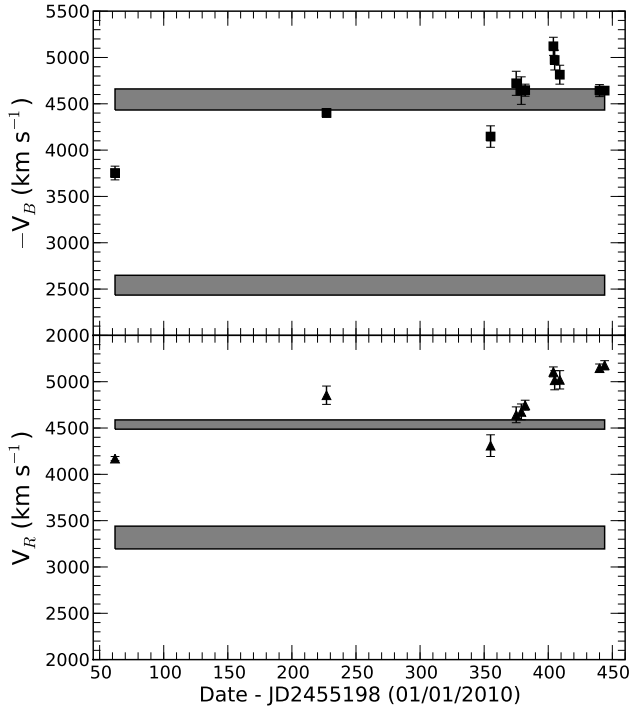}
  \caption{Evolution of the peak velocities of the blue (top panel)
    and red (bottom panel) peaks from 2010 March to 2011 March. The
    gray horizontal bars show the minimum and maximum values (plus
    uncertainties) in the spectra from 1991 to 2001 (SB03).}
  \label{vels}
\end{figure}

From $\lambda_{B}$ and $\lambda_{R}$ we obtained the blue and red
peak velocities $V_{B}$ and $V_{R}$ relative to the systemic
velocity of the galaxy, adopted as the one corresponding to the narrow
component of the H$\alpha$ emission, as we have done in our previous
studies (SB03). Figure \ref{vels} shows the temporal evolution
of $V_{B}$ and $V_{R}$. The velocities of the blue peak range
between $-5500$ and $-3500$ km s$^{-1}$ while for the red peak
they are between 4000 and 6000\,km\,s$^{-1}$.  A comparison between
these peak velocities with those from the period 1991--2001
(SB03) shows that, in the new observations, the blue and red
peaks always have larger peak velocities than the minimum values
observed between 1991 and 2001, which are $\approx
-2500$\,km\,s$^{-1}$ for the blue and $\approx 3250$\,km\,s$^{-1}$ for
the red, as indicated by the lower gray horizontal bars in both panels
of Figure\,\ref{vels}. Most velocities are instead similar or higher
than the maximum values observed between 1991 and 2001,
$\approx-4500$\,km\,s$^{-1}$ for the blue and $\approx
4500$\,km\,s$^{-1}$ for the red peak. These are indicated by the upper
gray horizontal bars in Figure\,\ref{vels}.
The values of V$_B$ and V$_R$ were obtained by using 
the expression for the relativistic Doppler effect.

\begin{figure}
 \centering
  \includegraphics[scale=0.55]{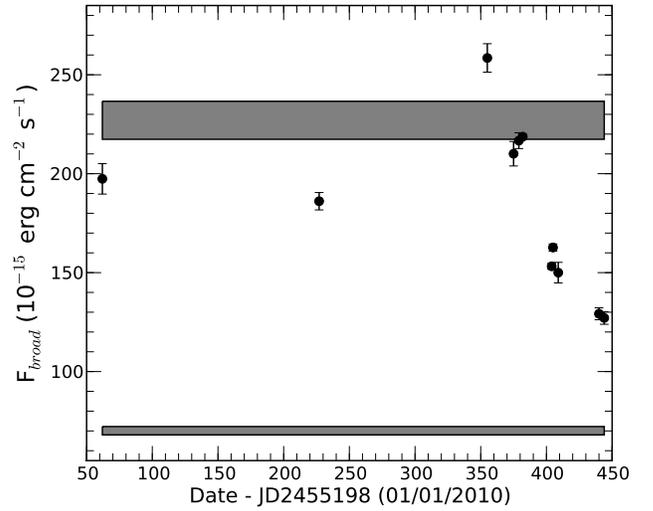}
  \caption{Evolution of the integrated flux of the broad double-peaked
    H$\alpha$ line $F_{\text{broad}}$ from 2010 March to 2011 March. The gray
    horizontal bars show the minimum and maximum values (plus
    uncertainties) in the spectra from 1991 to 2001 (SB03).}
  \label{flarga}
\end{figure}

Figure \ref{flarga} shows the variation of the integrated broad
H$\alpha$ emission line flux from 2010 March to 2011 March. The
integrated flux is always above the minimum value observed in the
period 1991--2003 (lower gray bar in the figure), being close to the
maximum value from 2010 March to 2011 January (upper gray bar in the
figure). A maximum in the flux occurred between 2010 August 15
and 2010 December 22, when the integrated flux rose above the maximum
value observed between 1991 and 2001. After 2010 December 23, there
was an abrupt decrease in the broad-line flux, by $\approx$\,25\%
of its maximum value in 20 days. On the other hand, it can be observed
that there is little variation on timescales shorter than a week (the
maximum time interval between the observations obtained in the same
month). Thus significant variations in the broad-line flux occur on a
timescale just above a week. From Figure \ref{flarga}, one can derive a
decrease of $\approx$\,12\% of the flux in 10 days.

\begin{figure}
 \centering
  \includegraphics[scale=0.56]{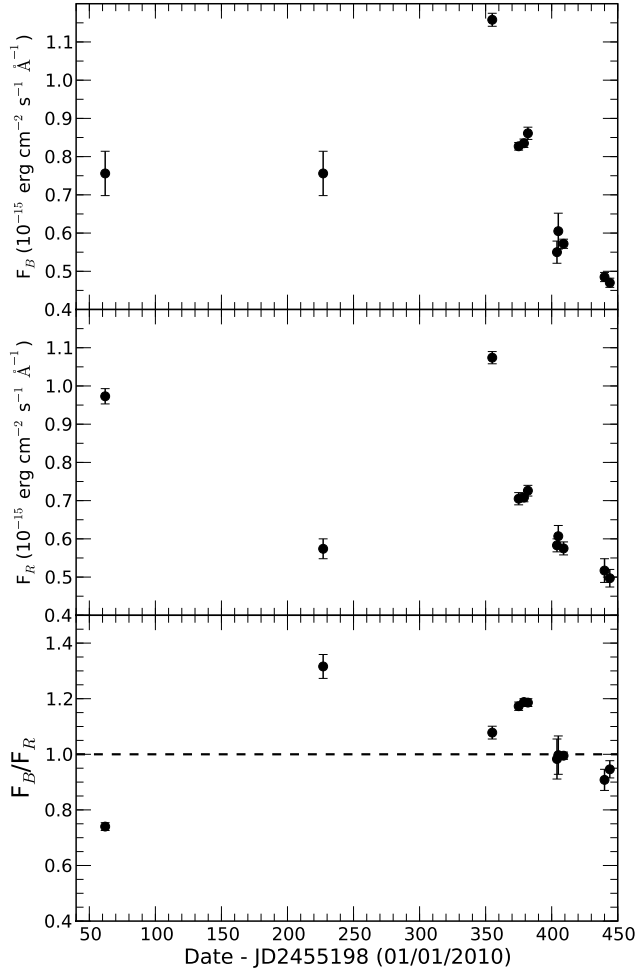}
  \caption{Evolution of the peak flux densities of the blue peak (top panel), 
    the red peak (middle panel) and of the ratio between the two
    (bottom panel) from 2010 March to 2011 March.}
  \label{fbfr}
\end{figure}

Figure \ref{fbfr} shows the temporal evolution of the peak fluxes of
the blue and red peaks as well as that of their ratio
$F_{B}$/$F_{R}$. On 2010 March, the red peak was stronger than the
blue, while in 2010 August the blue peak was stronger. Thus,
within a time interval of 5 months, there has been an inversion of the
$F_{B}$/$F_{R}$ ratio, and, in the following 7 months this ratio
decreased to a value which is just below 1, while in 2010 March this
value was 0.7. Thus the change from a stronger red peak to a
stronger blue peak was faster than the change back to a stronger
red peak. From Figure\,\ref{fbfr} we thus derive a timescale for
the inversion of the relative intensity of the two peaks of
$\approx$5 months.

\begin{figure}
\centering
\includegraphics[scale=0.45]{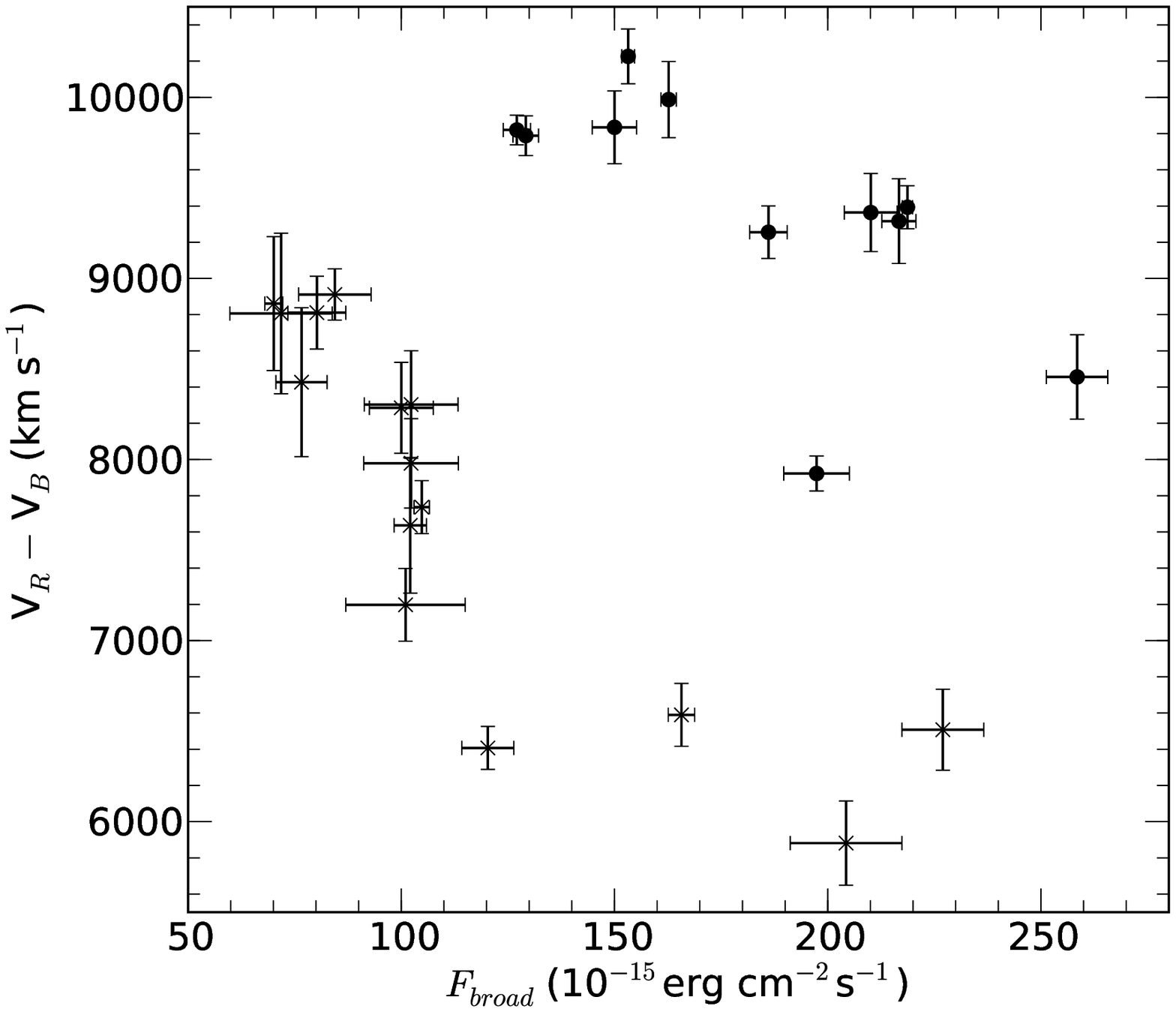}
\caption{The inverse correlation between $F_{\text{broad}}$ and $V_{R}$ $-$
  $V_{B}$. Circles represent the measures for this work and crosses
  represent the data from SB03. The inverse correlation for
  the new observations however is displaced to larger velocities and
  fluxes relative to previous observations (SB03). This suggests
  that emission from the inner parts of the disk became more important
  relative to that of the outer parts.}
\label{correlacao}
\end{figure}

Figure \ref{correlacao} shows that there is an inverse correlation
between the velocity separation of the red and blue peaks $V_{R}$ --
$V_{B}$ and $F_{\text{broad}}$, observed both in our recent observations,
represented by the circles, as well as in the previous observations of
SB03, represented by the crosses. This inverse correlation between
$F_{\text{broad}}$ and $V_{R} -$ $V_{B}$ can be interpreted as due to:
(1) when the central ionizing source becomes brighter, the line emission from the outer
radii become relatively more important,
as the radiation of the ionizing source reaches larger distances in the disk,
where the velocity is lower; (2) when the central source is fainter, 
the inner radii are favored, as the ionizing radiation reaches smaller radii,
where the velocity is larger. 

A comparison of the two data sets in Figure\,\ref{correlacao} shows that our recent data define a
sequence parallel to that of the previous data, displaced from the
previous data due to larger velocity separations between the
two peaks and to higher fluxes.  The larger velocity
separation of the two peaks suggests that the line emitting portion of
the disk is closer to the ionizing source or has a larger contribution from
gas closer to the source.

\begin{deluxetable*}{l c c c c c}
\tablecolumns{6}
\tablecaption{Measurements of the Profile Properties \label{medidas}}
\tablehead{   			&$\lambda_{B}$\tablenotemark{a}		&$\lambda_{R}$\tablenotemark{a}		&$F_{B}$\tablenotemark{b}					&$F_{R}$\tablenotemark{b}					&$F_{\text{broad}}$\tablenotemark{c}\\
Date				&(\AA)			&(\AA)			&(10$^{-15}$erg cm$^{-2}$ s$^{-1}$\AA$^{-1}$)	&(10$^{-15}$ erg cm$^{-2}$ s$^{-1}$\AA$^{-1}$)	&(10$^{-15}$ erg cm$^{-2}$ s$^{-1}$)}

\startdata
2010 Mar 4		&6483.0$\pm$1.6		&6656.5$\pm$0.5		&0.760$\pm$0.06					&0.970$\pm$0.02					&197.4$\pm$7.7\\
2010 Aug 15		&6469.0$\pm$1.0   	&6671.7$\pm$2.2   	&0.760$\pm$0.06  				&0.570$\pm$0.03   				&186.1$\pm$4.4\\
2010 Dec 22		&6474.5$\pm$2.5   	&6659.6$\pm$2.6   	&1.160$\pm$0.02   				&1.070$\pm$0.02   				&258.5$\pm$7.2\\
2011 Jan 11		&6462.1$\pm$2.8		&6667.0$\pm$1.9		&0.830$\pm$0.01					&0.700$\pm$0.02					&210.1$\pm$6.2\\   
2011 Jan 15		&6463.8$\pm$3.2		&6667.7$\pm$1.9   	&0.840$\pm$0.01   				&0.710$\pm$0.01   				&216.7$\pm$4.0\\   
2011 Jan 18		&6463.7$\pm$1.4   	&6669.3$\pm$1.2   	&0.860$\pm$0.02   				&0.730$\pm$0.01   				&218.7$\pm$1.2\\   
2011 Feb 8		&6453.5$\pm$2.1   	&6677.3$\pm$1.2   	&0.550$\pm$0.03   				&0.580$\pm$0.02   				&153.2$\pm$1.5\\   
2011 Feb 10		&6456.7$\pm$2.3   	&6675.3$\pm$2.3   	&0.600$\pm$0.05   				&0.600$\pm$0.03   				&162.7$\pm$1.8\\   
2011 Feb 14		&6460.1$\pm$2.2   	&6675.4$\pm$2.2   	&0.570$\pm$0.01   				&0.580$\pm$0.02   				&150.0$\pm$5.2\\   
2011 Mar 17		&6463.8$\pm$1.4   	&6678.2$\pm$1.0   	&0.490$\pm$0.01   				&0.520$\pm$0.03   				&129.2$\pm$3.0\\
2011 Mar 21		&6463.8$\pm$0.7   	&6678.9$\pm$1.1   	&0.470$\pm$0.01   				&0.500$\pm$0.02   				&127.1$\pm$3.2
\enddata

\tablenotetext{a}{$\lambda_{B}$ and $\lambda_{R}$ are the peak wavelength of the blue and red peaks respectively.}
\tablenotetext{b}{$F_{B}$ and $F_{R}$ are the corresponding peak fluxes.}
\tablenotetext{c}{$F_{\text{broad}}$ is the integrated flux of the broad double-peaked H$\alpha$ emission line.}

\end{deluxetable*}

\section{Accretion disk models}

In order to investigate the origin of the observed variations in the
double-peaked profile, we have used accretion disk models to reproduce
the profile. Figure \ref{fbfr} shows that the double-peaked profile
still shows variations in the relative intensities of the two peaks,
alternating between a stronger blue peak and a stronger red peak, as
previously observed by SB03. This pattern of variation
supports the origin of the profile in an accretion disk whose emission is
not axisymmetric. In SB03, the relative strengths of the two peaks and their variation
were better explained by emission from a circular accretion disk with a single spiral arm. The enhanced
emission in the arm, as it rotates in the disk, produces the observed
asymmetry of the profile. We will thus also adopt in our present modeling the
single spiral arm model to fit the double-peaked profiles.

The source of radiation which ionizes the disk in order to drive the
H$\alpha$ emission is also relevant to the modeling. As pointed out in the
Introduction, in the case of NGC\,1097, this source is an RIAF
internal to the line-emitting part of the disk \citep{Nemmen06}. 
\citet{Dumont90} have found that, in the case of an accretion
disk illuminated by a non-thermal continuum source with a spectral
energy distribution extending up to the hard X-ray band, there is a
``radius of maximum emission'' or ``saturation radius'', at which the
H$\alpha$ emission reaches a maximum value. Thus we consider an additional model in which the disk emissivity may
increase with radius up to the saturation radius and then decrease
outward.

With the above considerations in mind, we then fitted the
double-peaked H$\alpha$ line profiles with two different models,
described below. In both models we assume a circular emitting disk, or
wide ring (as it has a hole in the middle), with a perturbation in the
emissivity in the form a spiral arm. In the first model which we call \textit{Spiral Arm Model}
the variations in the relative strengths of the peaks are controlled
by the rotation of the spiral arm while the separation of the peaks
is controlled by an emissivity law which is a simple power-law function
of the radius.  By changing the power-law index we can weight the
emissivity of the disk toward inner regions (leading to a broader
profile) or the outer regions (leading to a narrower profile). In the second model which we call the
\textit{Saturated Spiral Model} we have changed the emissivity law to a broken
power-law, which can produce a ring of maximum emissivity between the
inner and outer radii. In this case, the separation of the peaks is
controlled by the radius of this ring of maximum emissivity.

\subsection{Spiral Arm Model}

We adopt the formulation described in \citet{Gilbert99}, in which: (1)
the line emitting portion of the disk is circular and located between
an inner radius $\xi_{1}$ and an outer radius $\xi_{2}$ (where $\xi$
is the disk radius in units of the gravitational radius
$r_g={GM_{\bullet}}/{c^2}$, $c$ is the light speed, $G$ is the gravitational constant and
$M_{\bullet}$ is the mass of the black hole); (2) the disk has an inclination $i$
relative to the line of sight (zero degrees is face on). Superimposed
on the axisymmetric emissivity of the circular disk, there is a
perturbation in the form of a spiral arm. We adopt only one spiral arm, as 
we have already verified that two or more arms do not
reproduce well the variations in the profile (SB03).

The total emissivity of the disk is given by:

\begin{multline} \label{arm}
\epsilon(\xi,\phi) = \epsilon (\xi) \left \{ 1 + \frac{A}{2} \exp \left [ -\frac{4 \ln2}{\delta^{2}} (\phi - \psi_{0})^{2} \right ] \right . \\ 
\left . + \frac{A}{2} \exp \left [ -\frac{4 \ln2}{\delta^{2}} (2\pi - \phi + \psi_{0})^{2} \right ] \right \},
\end{multline}

where 
\begin{equation}
 \epsilon(\xi) = \epsilon_{0}\xi^{-q}
\label{epsilonzero}
\end{equation}

\noindent is the axisymmetric emissivity of the disk, $A$ is the
brightness contrast between the spiral arm and the underlying disk,
and the expression between square brackets represents the decay of the
emissivity of the arm as a function of the azimuthal distance $\phi -
\psi_{0}$ from the ridge line to both sides of the arm, assumed to be
a Gaussian function with FWHM $\delta$ (azimuthal width).

The relation between the azimuthal angle $\phi_{0}$ and the angular
position $\psi_{0}$ of the ridge of emissivity on the spiral arm is given by

\begin{equation}
 \psi_{0} = \phi_{0} + \frac{\ln(\xi/\xi_{\text{sp}})}{\tan p},
\end{equation}
\noindent where $\phi_0$ is the azimuthal angle of the spiral pattern,
$p$ is the \textit{pitch} angle and $\xi_{\text{sp}}$ is the innermost radius
of the spiral arm.

The specific intensity from each location in the disk, in the frame of
the emitting particle is calculated as:

 \begin{equation}
  I(\xi,\phi,\nu_{e}) = \frac{\epsilon(\xi,\phi)}{4\pi}\frac{e^{-(\nu_{e}-\nu_{0})^{2}/{2\sigma^{2}}}}{(2\pi)^{1/2}\sigma},
 \end{equation}

\noindent where $\nu_{e}$ is the emission frequency, $\nu_{0}$
(H$\alpha$ 6564.6\AA) is the rest frequency and $\sigma$ is the local
``broadening parameter'' \citep{CeH89}.

In the fits of the individual profiles we first used the same set
of parameters of SB03, namely: the inner and outer radii of
the line emitting portion of the disk $\xi_{1} = \xi_{\text{sp}} = 450$ and $\xi_{2} = 1600$, 
the inclination angle $i =34^{\circ}$, broadening parameter $\sigma = 1200$ km s$^{-1}$,
 pitch angle $p=50^{\circ}$ and $\delta=70^{\circ}$. In order to reproduce
the evolution of the profile we varied the parameters $A$, $q$ and
$\phi_{0}$. The variation of $\phi_{0}$ regulates the evolution of the
shape of the profile, while a variation in $A$, together with $\phi_{0}$, regulates the relative
intensities of the two peaks. The value of $q$ regulates the radius
in the disk of the maximum line luminosity, which corresponds to the inner
regions for the broadest profiles, or to the outer regions for the
narrowest profiles.

 \begin{figure}
  \centering
    \includegraphics[scale=0.65]{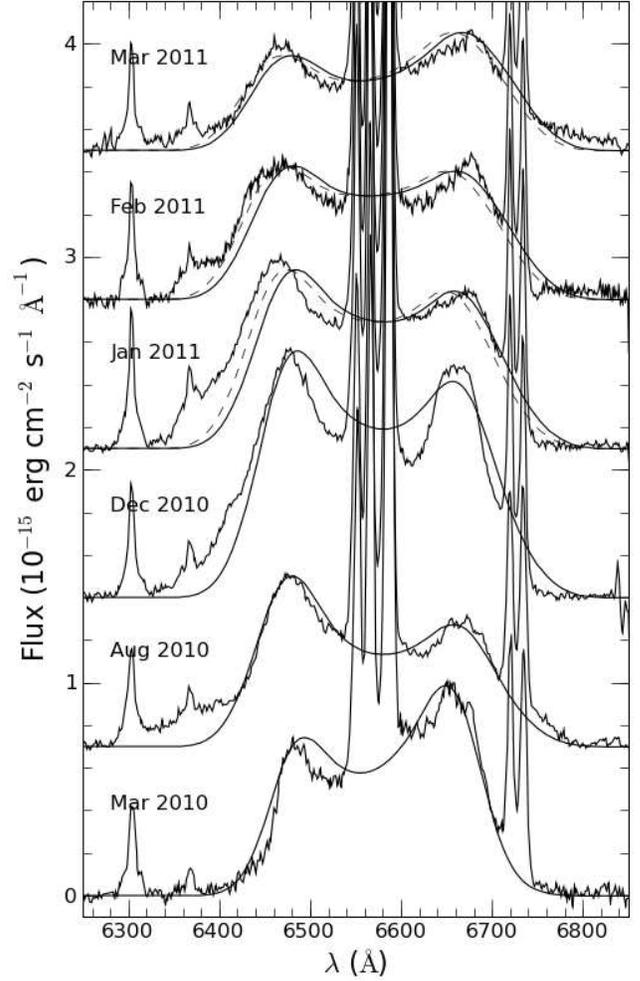}
    \caption{Best fits of the spiral arm model for the epochs from
      2010 March 4 to 2011 March 21, obtained by varying
      $\phi_{0}$ (orientation of the spiral pattern), $A$ (contrast
      between the spiral arm and the disk), and $q$ (the slope of the
      emissivity law). The values of the parameters are listed in
      Table \ref{parspiral}. }
    \label{mod1}
  \end{figure}

\begin{deluxetable}{l c c c c}
 \centering
  \tablecolumns{5}
  \tablecaption{Parameters for the Spiral Arm Model Fits. \label{parspiral}}
  \tablehead{  		&		&		&$\phi_{0}$	 	&Shift \\
	Date  		& $A$  		&$q$ 		&(deg)			&(km s$^{-1}$)}
  \startdata
2010 Mar 4		&2.0		&$-$2.0		&130			&0\\
2010 Aug 15		&1.0		&1.0		&250			&0\\
2010 Dec 22		&0.3		&1.0		&330			&0\\
2011 Jan 11		&0.5		&1.5		&340			&$-$460\\
2011 Jan 15		&0.5		&1.5		&340			&$-$460\\
2011 Jan 18		&0.5		&1.5		&340			&$-$460\\
2011 Feb 8		&1.0		&2.0		&360			&$-$460\\
2011 Feb 10		&1.0		&2.0		&360			&$-$460\\
2011 Feb 14		&1.0		&2.0		&360			&$-$460\\
2011 Mar 17		&2.0		&2.0		&400			&$-$460\\
2011 Mar 21		&2.0		&2.0		&400			&$-$460
\enddata
\end{deluxetable}

The best-fitting models are compared to the observed line profiles in
Figure \ref{mod1}, while the corresponding parameter values ($A$, $q$ and
$\phi_{0}$) are listed in Table \ref{parspiral}. The contrast of the
spiral arm $A$ ranged from 2.0 at the first epoch, decreased to a
minimum value of 0.5 in 2011 January, and increased again to 2.0 in
2011 March, while $q$ increased from the first to the last epochs. An
abrupt variation in $q$ occurred between 2010 March and 2010 August
when the value of $q$ changed from $-2.0$, favoring the outer parts of
the disk, to 1.0, favoring the inner parts of the disk. The $\phi_{0}$
values were constrained to provide a monotonic rotation of
the spiral pattern, from 130$^{\circ}$ to 400$^{\circ}$. The spiral
arm has thus almost completed one revolution in one year. 

For the spectra from 2011 January to 2011 March, an improvement of the fit can be obtained by
allowing a small, ad hoc blueshift of the central wavelength of
the profile relative to the wavelength of the narrow H$\alpha$ line
(6564.6\AA). This blueshift is $\approx$ $-460$
km\,s$^{-1}$, as listed in the last column of Table
\ref{parspiral}. The improved fits with the
blueshifts are shown as dashed lines in Figure \ref{mod1}.

\begin{figure}
 \centering
  \includegraphics[scale=0.45]{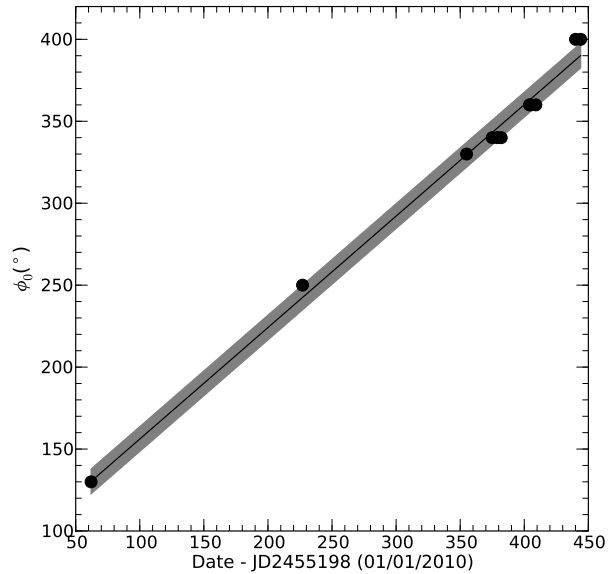}
  \caption{Evolution of the $\phi_{0}$ values obtained from the fit of
    the spiral arm model. The solid line is a linear fit to the data
    and the shaded bar covers the uncertainty in the fit.}
  \label{phi1}
\end{figure}

In Figure \ref{phi1} we show the evolution of the $\phi_{0}$ values
obtained from the fits, represented by the circles in the figure,
while the black solid line is the best linear regression to the points
and the gray band covers the uncertainty in the regression. The linear
regression has angular coefficient $\dot{\phi_{0}} = 0.680\pm\fdg0.02$ day$^{-1}$, which represents the angular velocity of the arm and
implies a rotation period of $P\,\approx 17.7\,\pm$0.5 months.

\subsection{Saturated Spiral Model}

In an effort to improve the fit to the broad double-peaked H$\alpha$ profiles,
we modified the spiral arm model to allow a more flexible radial emissivity
law. We have chosen an emissivity law which allows the existence
of a ``radius of maximum emissivity'' between the inner and outer radii of the disk at
which the broad H$\alpha$ line reaches a maximum intensity, as
proposed by \citet{Dumont90}. The emissivity increases with radius until the saturation radius
and then decreases outward.

The radius of maximum emissivity was incorporated by changing the
axisymmetric part of the emissivity of the disk
(Equation \ref{epsilonzero}) as follows:
\begin{equation}
 \epsilon(\xi) =\left\{ \begin{array}{ll} \epsilon_{0}\xi^{-q_{1}} & ,\,\xi_{1} < \xi < \xi_{q}\\ \epsilon_{0}{\xi_{q}}^{-(q_{1}-q_{2})}\xi^{-q_{2}} &,\,\xi_{q} < \xi < \xi_{2} \end{array} \right.
\end{equation}

\noindent where: the new parameter $\xi_{q}$ is the radius of maximum
emissivity, or saturation radius, at which the emissivity law changes;
$q_{1}$ is the index of the emissivity law for the regions internal to
$\xi_{q}$ ($\xi_{1} < \xi < \xi_{q}$);  $q_{2}$ is the index for the
regions external to $\xi_{q}$ ($\xi_{q} < \xi < \xi_{2}$);
$\xi_{1}$ and $\xi_{2}$ are the inner and outer radii, respectively.

We kept the value $q_{2}=3.0$, as discussed by \citet{Dumont90}, who
showed that beyond the saturation radius the flux emitted by the line
decreases as $\propto\,\xi^{-3}$. After many tests, $q_{1}=-2.0$ gave
the best fits, and was fixed at this value. The negative value allows
the increase of the emissivity until the saturation radius. The
parameter $\xi_{q}$ was allowed to vary.  We kept the parameters of
the spiral arm at the values $\xi_{\text{sp}}=\xi_{1}$, $p=50^{\circ}$,
$\delta=70^{\circ}$, as well as the inclination of the disk at
$i=34^{\circ}$.

We have varied the broadening parameter and found that the best value
to reproduce the shape of most profiles was $\sigma = 900$ km
s$^{-1}$.

In summary, in the Saturated Spiral model, the only parameters we needed to vary in
order to reproduce the evolution of the profile were $\xi_{q}$, $A$,
and $\phi_{0}$. The changes in the separation between the blue and red
peaks were obtained by changing $\xi_{q}$; when $\xi_{q}$ is smaller,
the separation between the blue and red peaks increases.

  \begin{figure}
   \centering
    \includegraphics[scale=0.65]{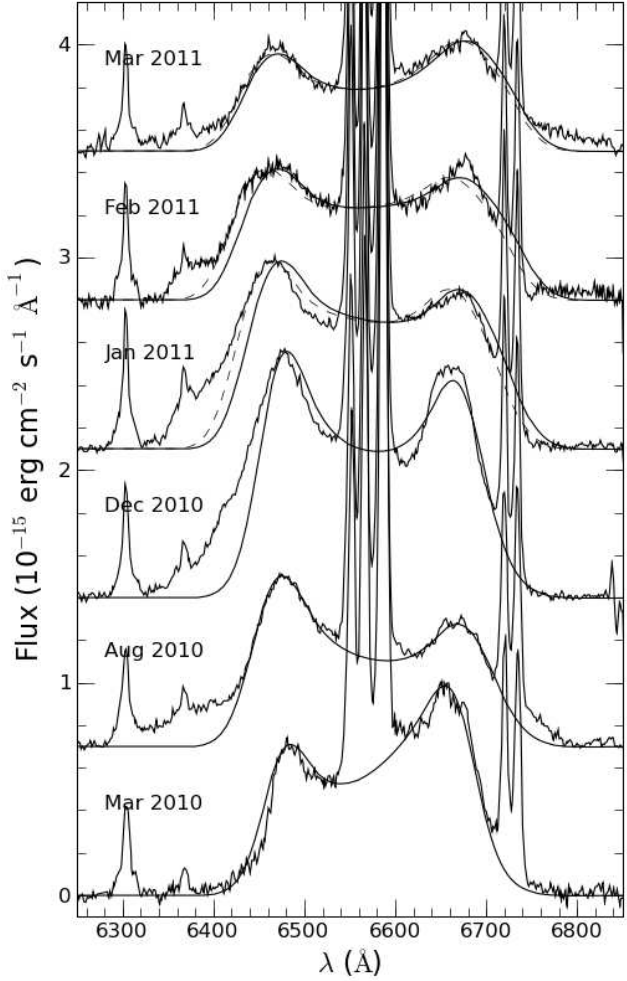}
    \caption{Best fits for the hybrid model, obtained by varying
      $\phi_{0}$ (orientation of the spiral pattern), $A$ (contrast
      between the disk and the spiral arm) and $\xi_{q}$ (radius of
      maximum emissivity). The values of the parameters are listed
      in Table \ref{parshibrid}. }
    \label{mod3}
  \end{figure}

\begin{deluxetable}{l c c c c}
 \centering
  \tablecolumns{5}
  \tablecaption{Parametes for the Saturated Spiral Model Fits \label{parshibrid}}
  \tablehead{ 		& 		& 			&$\phi_{0}$		 	&Shift \\
Date			&$A$ 		&$\xi_{q}$ 		&(deg)				& (km s$^{-1}$)}
  \startdata
2010 Mar 4		&3.0		&1200			&140				&0\\
2010 Aug 15		&1.0		&800			&220				&0\\
2010 Dec 22		&0.2		&1000			&330				&0\\
2011 Jan 11		&1.0		&650			&340				&$-$500\\
2011 Jan 15		&1.0		&650			&340				&$-$500\\
2011 Jan 18		&1.0		&650			&340				&$-$500\\
2011 Feb 8		&1.0		&550			&360				&$-$550\\
2011 Feb 10		&1.0		&550			&360				&$-$550\\
2011 Feb 14		&1.0		&550			&360				&$-$550\\
2011 Mar 17		&1.5		&600			&400				&$-$230\\
2011 Mar 21		&1.5		&600			&400				&$-$230
\enddata
\end{deluxetable}

The best-fitting models are compared with the observed line profiles
in Figure \ref{mod3}, and the corresponding parameters values
($\xi_{q}$, $A$ and $\phi_{0}$) are listed in
Table \ref{parshibrid}. The parameter $\xi_{q}$ ranged from a maximum
value of 1200 in 2010 March when the separation between the
peaks was the smallest, to a minimum value of 550--600 in
2011 February--March, when the separation between the peaks was the
largest. The parameter $A$ assumed the value 3.0 in 2010 March, when
the red peak was significantly stronger than the blue, as seen in
Figure \ref{mod3}, while in 2010 December this parameter reached the
minimum value of 0.2, when the peaks had similar strengths. Similarly
to the previous model, $\phi_{0}$ varied from 140$^{\circ}$ in the
first epoch to 400$^{\circ}$ in the last.

As in the case of the Spiral Arm model, from 2011 January to 2011 March,
the fit was improved by introducing a blueshift  which is listed in the last column of 
Table \ref{parshibrid}. This blueshift ranges from $\approx$ $-230$ to $-550$
km\,s$^{-1}$, and the improved fits with the
blueshifts are shown as dashed lines in Figure \ref{mod3}.

In Figure \ref{modelo2} we show emissivity maps constructed using the parameters of the 
best fits using the Saturated Spiral model. In 2010 March the radius of
maximum emissivity $\xi_q$ was closer to $\xi_{2}$, when the outer parts where
brighter, while in 2011 March $\xi_q$ was closer to
$\xi_{1}$ and the inner parts were brighter.

\section{Discussion}

 \subsection{Timescales}
In order to discuss the physical processes driving the changes in the
profiles, we first review four relevant timescales for the accretion
disk in NGC\,1097: the \textit{viscous timescale} ($\tau_{\text{visc}}$), the  \textit{sound-crossing timescale} ($\tau_{s}$), the \textit{dynamical timescale} ($\tau_{\text{dyn}}$) and the \textit{light travel timescale} ($\tau_{l}$).
These timescale are presented in \citet{Frank} and can be modified in order to express the disk radius in
terms of the gravitational radius \citep{Eracleous98, Lewis10}:

\begin{equation}
 \tau_{\text{visc}} = 10^{6} M^{3/2}_{8} \xi^{5/4}_{3} \alpha^{-4/5}_{-1} \dot{M}^{-3/10}_{-1}\, \text{years}
\end{equation}

\begin{equation}
 \tau_{s} = 70 M_{8} \xi_{3} T_{5}^{-1/2}\,\text{years}
\end{equation}

\begin{equation}
 \tau_{\text{dyn}} = 6M_{8}{\xi_{3}}^{3/2}\,\text{months}
\end{equation}

\begin{equation}
 \tau_{l} = 6 M_{8}\xi_{3}\,\text{days}
\end{equation}

\noindent where $\xi_{3}=10^{-3}\xi$, M$_{8}=M_{\bullet}\times10^{-8}{M_{\odot}}^{-1}$,
 $\alpha=10^{-1}\times\alpha_{-1}$ is the Shakura--Sunyaev viscosity parameter, $T_{5}$
is the temperature in units of 10$^{5}$ K, and
$\dot{M}_{-1}$ is the mass accretion rate in units of 0.1 $M_{\odot}$\,yr$^{-1}$.

For NGC 1097, $M_{\bullet} = 1.2\,\times\,10^8M_{\odot}$ \citep{Lewis}, and 
$\dot{M}/\dot{M}_{\text{Edd}} = 0.0064$ (in Eddington units, \citet{Nemmen06}), corresponding to
$\dot{M}=0.018$ M$_{\odot}$\,yr$^{-1}$. Adopting typical values of $\alpha=0.1$ and
T$_{5}= 0.1$, a characteristic radius for
the disk is $\xi=1000$. Adopting the above values of $M_{\bullet}$ and $\dot{M}$, we
obtain the following estimates for the timescales:

\begin{enumerate}
 \item {$\tau_{\text{visc}} \gtrsim$ 10$^7$ years}: this large value for the viscous timescale does not support that the observed changes in the profile---on timescales from weeks to years---are due to inward transfer of mass due to viscous torques;

 \item {$\tau_{s} \sim$ 265 years}: this is also much larger than the observed variations timescales;

 \item {$\tau_{\text{dyn}} \gtrsim 7$ months}: the dynamical timescale depends on the radius, with the inner parts of the disk having
shorter timescales than the outer parts. Adopting $\xi_{3}
=0.45$ for the inner radius, as obtained from our modeling, the
corresponding dynamical timescale is just above 2 months. The new
observations presented here show that the variation timescale of
the relative strengths of the blue and red peaks ($F_{B}$/$F_{R}$) is
$\approx$\,5 months (see Figure \ref{fbfr}), thus compatible with the dynamical time.   
In our modeling, this variation was
reproduced by the rotation of the spiral pattern around the disk, whose
period is close to the dynamical time at $\xi_2$;

 \item{$\tau_{l} \gtrsim 7$ days}: considering the range of radii of the disk,
this timescale ranges from $\approx$4 to $\approx$12 days. This is the
shortest timescale of the disk. According to our data (see Figure \ref{flarga}), the smallest timescale
of the profile variations---namely variations in the integrated flux
of the double-peaked line---occur in $\le$7 days. This supports the
interpretation that the variability in the integrated flux $F_{\text{broad}}$
occurs in the light travel timescale, as a result of variations in the
ionizing flux. Similarly, the variations in $V_{B}$ and $V_{R}$---see
Figure \ref{vels}---which correlate with $F_{\text{broad}}$, also occur in the
light travel timescale. According to the models, these are due to
changes in the region of maximum emissivity of the disk.
\end{enumerate}

\subsection{Constraints on the physical mechanisms causing the variations}

The timescales discussed above provide constraints on the mechanisms
that cause the changes in the double-peaked profile. Our data clearly
reveal two distinct timescales: (1) the first corresponding to
changes in the integrated flux of the double-peaked profile, which is
just over 7 days; and (2) the second corresponding to the changes in
relative strengths of the blue and red peaks, which is in the range
5--6 months.

We have identified the first timescale with the \textit{light travel
  timescale}. In order to model the corresponding variations in the
flux of the double-peaked line, and at the same time reproduce the
variation of the width of the profiles it was necessary to vary the
disk emissivity. When the flux is higher, the profile is narrower, and
the emissivity is higher in the outer parts of the disk, while when
the flux is lower, the profile is broader and the emissivity is
relatively higher in the inner parts of the disk. In the Saturated Spiral model,
what varies is the radius of maximum emissivity.

The physical scenario we favor is the following: the double-peaked
profile originates in gas rotating in a thin accretion disk, which is
ionized by an RIAF \citep{Nemmen06} located inside the inner radius of
the line-emitting portion of the disk. As the variations occur
on the \textit{light travel timescale}, our interpretation is that
the variation in $F_{\text{broad}}$ can be considered a ``reverberation''
effect produced by variations in the ionizing flux from the central
RIAF, and allows a test for the values of the radii obtained via the
models. Calculating the values for the inner and outer radii of the
disk in light days we obtain for $\xi_1$\,$\approx$\,3 light days and
for $\xi_2$\,$\approx$\,11 light days, which is consistent with 
the variability timescale of $\approx$\,7 days that we observe.

In order to further confirm the reverberation hypothesis, it would be
necessary to do a proper ``reverberation mapping campaign''
(e.g., \citet{Peterson04}) of the accretion disk via simultaneous
observations of the RIAF emission (e.g., in X-rays) and the 
double-peaked profile. We plan to pursue such a monitoring campaign in the future,
 including a more extended spectral coverage to include the blue part of the optical 
spectrum which will also allow us to test other models for the origin of the continuum, 
such as the one proposed by \citet{Gaskell11}.

We have identified the second timescale---of the variations in the
relative strengths of the blue and red peaks---with the
\textit{dynamical timescale}. This means that the asymmetry in the
profile is due to a non-axisymmetric emitting structure -- modeled as
a single spiral arm -- rotating in the disk, with a rotation period of
$\approx$\,18 months. This timescale and its interpretation differs
from our previous one (SB03). In that paper we concluded that
the variation of the relative strengths of the two peaks occurred in
the \textit{sound-crossing timescale}. This distinct interpretation
was due to the fact that the period we have derived then for the
variation was 5.5\,yr, and at that time our estimate for the SMBH
mass was $\approx$\,10$^6$ $M_\odot$, giving a smaller value for the
sound-crossing timescale, consistent with the derived period of
variation. With a revised SMBH mass two orders of magnitude larger,
the revised sound-crossing timescale
is much larger than the timescale of the variation of the
relative fluxes of the blue and red peaks we have now derived 
from our more frequent observations.

\begin{figure*}{SATURATED SPIRAL MODEL}
\centering
\includegraphics[scale=0.6]{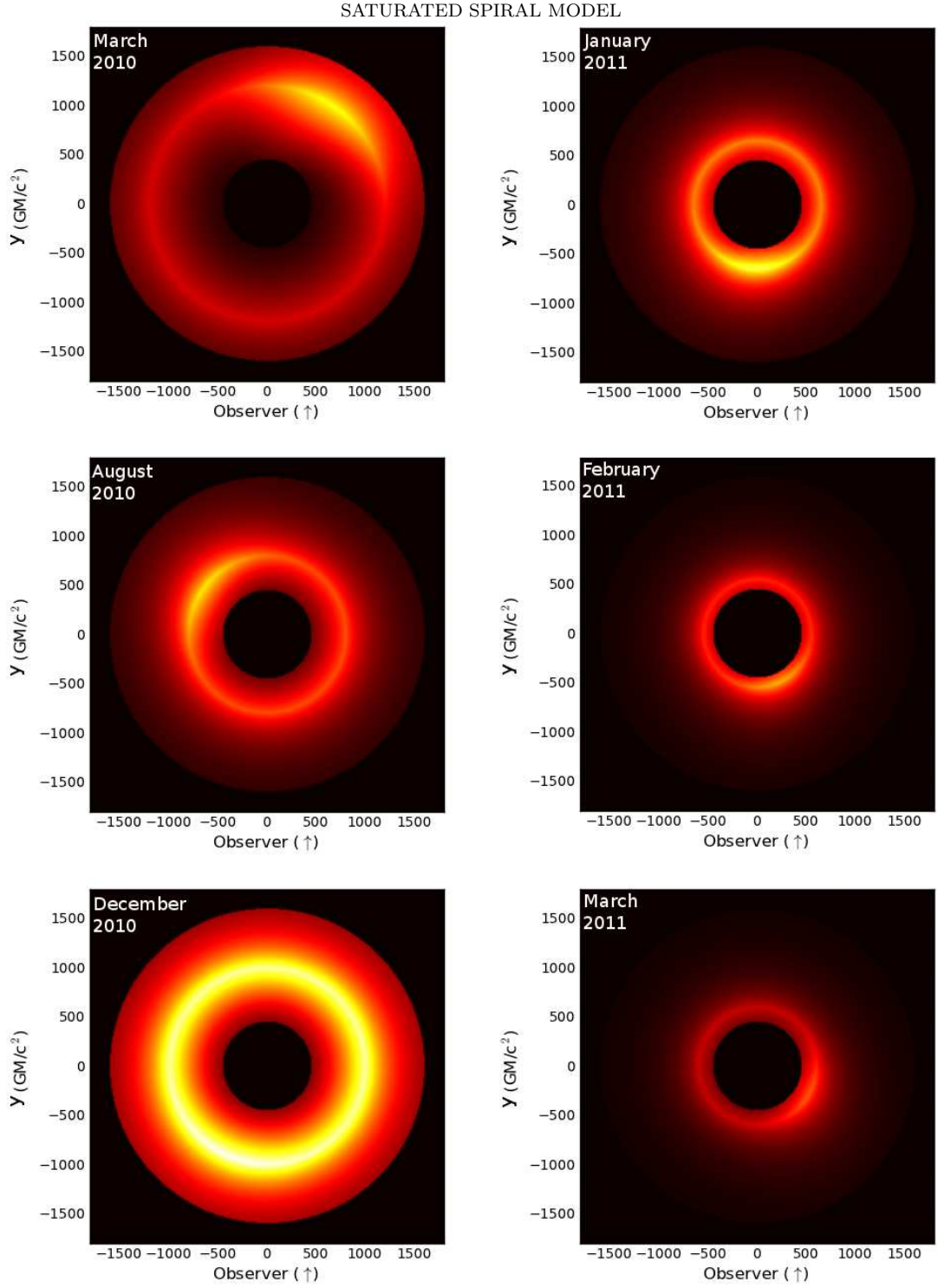}
\caption{Images showing the disk emissivity from 2010 March to 2011 March, using the hybrid model. White represents the brightest
  regions, and the observer is to the bottom. The disk parameters for
  these epochs are listed in Table \ref{parshibrid}.  }
\label{modelo2}
\end{figure*}

Is important to point out that the perturbation does not need to be a spiral arm, even though we
used the spiral arm prescription to model it. In the images of the disk, constructed using
the best model parameters for the different epochs, shown in Figure \ref{modelo2},
the spiral arm actually looks more like an azimuthal arc. Therefore, it could be that 
this arc is actually a clump that was produce by self-gravity and then sheared (see discussion in \citet{Flohic}).
Another alternative is that this feature is an irradiation-induced warp \citep{Pringle, Maloney}. But although the
irradiation-induced warp model can reproduce the observed variability of the line profiles \citep{Wu},
the required rotation timescale is incompatible with the mass of the SMBH in NGC 1097 (see Equation (4) in \citet{SB97}). On the other hand, the sheared clump would orbit the SMBH on the dynamical timescale and therefore could reproduce the 5 month variation that we observe.

Figure\,\ref{correlacao} shows that the profiles got broader and brighter in the recent
(2010--2011) observations when compared to the previous ones (1991-2001), although
at both epochs there is an inverse correlation between the flux and width
of the profiles. This inverse correlation is expected in the
reverberation scenario discussed above, but there is an offset between
the previous and present correlations. A comparison between the
parameters of the present and previous spiral arm model
shows that the emissivity law in the recent epochs favors a higher
contribution of the inner regions relative to the outer regions of the
disk, leading to a broader profile. One possibility to explain this
increase in emissivity of the innermost regions is an increase in the gas density in these
regions of the disk in the most recent epochs when compared to the period 1991--2001.

A comparison between Figures \ref{mod1} and \ref{mod3} shows that the Saturated Spiral model
gives a better fit to the double-peaked profiles. In order to quantify the goodness of the fits, we have 
obtained the root mean square (rms) deviation between the observed profiles and the models.
In almost all observations the Saturated Spiral model provided a smaller rms
when compared with that for the Single Spiral arm model, confirming what is seen in the fits.

Finally, we point out that, although the models can in general
reproduce the H$\alpha$ profiles, there are several epochs in which we
observe an excess flux in the blue wings of the profiles relative to
the models. This excess flux is observed from 2010 August to February
2011 (Figures \ref{mod1} and \ref{mod3}). In addition, from 2011 January to
March we observe a systematic blueshift of the center of the
line, from $-$230 to $-$550 km s$^{-1}$ which could be due to an accretion
disk wind. Therefore, it will be worthwhile in future work to compare
the observed line profiles with models that include an outflowing wind
(e.g., \citet{Hall})

\subsection{The mass of the supermassive black hole}

The physical value of the disk timescales, such as 
the light-travel timescale ($\tau_{l}$), as well as the inner and outer radii of the accretion disk,
are strongly dependent on the mass of the SMBH. Thus, uncertainties
in its determination must be taken into account. We have adopted here the SMBH mass value of
\citet{Lewis}: $M_{\bullet} = 1.2 \pm 0.2 \times 10^{8}$ $M_{\odot}$. 

The range of values for the inner and outer radii of the accretion disk ($\xi_1=450$ and $\xi_2=1600$, respectively) resulting from the uncertainties in the SMBH mass quoted by \citet{Lewis} are $2.7 \le \xi_{1} \le 3.6$ light-days and  $9.1 \le \xi_{2} \le 12.8$ light-days. But \citet{Lewis} obtain the value for the SMBH mass using the $M_{\bullet}$---$\sigma$ relation of \citet{Tremaine}, pointing out that they do not take into account the scatter in this relation. They consider only the uncertainty in the coefficients of the relation that results from the scatter (effectively they assume that the observed scatter is not intrinsic). If we now consider that the uncertainty in the SMBH mass is also affected by the scatter in the $M_{\bullet}$---$\sigma$ relation from \citet{Tremaine}, which can introduce an uncertainty in the mass of $\approx$ 0.3\,dex, the resulting minimum and maximum values for the inner and outer radii become: (1) for {$M_{\bullet} \approx 0.6 \times 10^{8}$ $M_{\odot}$}, the minimum radii are $\xi_{1} \approx 1.5 $ light-days and  $\xi_{2} \approx 5.5$ light-days; (2) for {$M_{\bullet} \approx 2.4 \times 10^{8}$ $M_{\odot}$}, the maximum radii are $\xi_{1} \approx 6 $ light-days and  $\xi_{2} \approx 22$ light-days.

Thus, even considering the maximum uncertainties which could affect the determination of the SMBH mass, the resulting values for the inner and outer radii of the disk are still of the order of light-days.  The above results support the interpretation that  the shortest variation timescale of 7\,days is the light-travel time between the ionizing source at the center of the disk and the mean radius of the disk. In fact, this observed timescale  can actually constrain the values of $\xi_{1}$ to $\le\,7$\,light-days and of ${\xi}_2$ to $\ge\,7$\,light-days. 

In addition, if we assume that 7\,light-days is the mean distance between the broad-line-emitting clouds and the SMBH,
 we can estimate the mass of the SMBH independently from the model, as follows. If this system of clouds is flattened,
 as suggested by their double-peaked profile, and we adopt as the corresponding Keplerian velocity around the SMBH the
 mean velocity of the peaks---$\approx 4650$\,km\,s$^{-1}$---we obtain 
$M_{\bullet}$\,=\,1.1\,$\times\,10^8$\,$M_\odot$, for an inclination of 34$^\circ$ relative to the plane 
of the sky  (as obtained from the disk model). This value is in good agreement with the one obtained by \citet{Lewis}.

\section{Conclusions}

We report the discovery of short timescale variations in the broad double-peaked
H$\alpha$ emission line from the LINER nucleus of NGC\,1097, observed in
9 epochs between 2010 March and 2011 March. They comprise: (1) variations in the
integrated flux and width of the line over a timescale of $\ge$\,7 
days, which we identified as the light-travel time between the
ionizing source and the disk; it is the first time that such short
timescale variations have been seen in a double-peaked profile
from the nucleus of a galaxy; (2) variations in the relative
intensity of the blue and red peaks on a timescale of 5--6 months,
which is compatible with the disk dynamical timescale.

Using two distinct accretion disk models we reproduced these variations through
the combination of two effects:

\begin{enumerate}
\item{short timescale changes in the disk emissivity---namely in the
contrast $A$ and index $q$ in the \textit{Spiral Arm model} and $A$
and $\xi_q$ in the \textit{Saturated Spiral model};}

\item{the rotation of the spiral arm---via the change of the
orientation $\phi_{0}$ of the arm in both models---resulting in a
estimate for the rotation period of the arm of $\approx 18$ months.}
\end{enumerate}

The values of the inner and outer radii, obtained from previous
modeling, were kept in the present modeling and are, respectively,
$\xi_1 =450$ and $\xi_2 =1600$ (corresponding to $\approx 3$ and 11
light-days, respectively). These values are in good agreement with the
short timescale variations seen in our observations. This agreement
supports the identification of the shortest timescale variations with
the light travel time between the ionizing source and the accretion
disk, in one hand, and gives further support to the model and the
derived parameters, on the other hand.


Our recent observations support the scenario in which an accretion
disk whose emissivity changes due to reverberation of a variable
ionizing continuum (presumably from an RIAF), has a non-axisymmetric
feature rotating in the dynamical timescale.

\acknowledgments{We thank the referee, Martin Gaskell, for the careful reading of the manuscript and his thoughtful comments.

Based on observations obtained at the Gemini Observatory, which is operated by the 
Association of Universities for Research in Astronomy, Inc., under a cooperative agreement 
with the NSF on behalf of the Gemini partnership: the National Science Foundation (United 
States), the Science and Technology Facilities Council (United Kingdom), the 
National Research Council (Canada), CONICYT (Chile), the Australian Research Council (Australia), 
Ministério da Ci\^encia, Tecnologia e Inova\c c\~ao (Brazil) 
and Ministerio de Ciencia, Tecnolog\'ia e Innovaci\'on Productiva (Argentina).
JSS and TSB acknowledge the Brazilian institutions CNPq, CAPES, and FAPERGS for partial support.}

\end{document}